\documentclass[twocolumn]{aastex63}

\received{September 8, 2020}
\revised{September 18, 2020}
\accepted{September 21, 2020}

\submitjournal{ApJS}

\shorttitle{Deep Space Network Digital Spectrometer}
\shortauthors{Virkler et al.}

\graphicspath{{./}{figures/}}

\begin{document}

\title{A Broadband Digital Spectrometer for the Deep Space Network\footnote{$\copyright$ 2020 All rights reserved.}\footnote{Released on September 2020}}

\correspondingauthor{Kristen Virkler}
\email{kristen.virkler@jpl.nasa.gov}

\author[0000-0002-3207-5117]{Kristen Virkler}
\affiliation{Jet Propulsion Laboratory, California Institute of Technology, 4800 Oak Grove Dr.,
  Pasadena, CA  91109, USA}

\author{Jonathon Kocz}
\affiliation{Department of Astronomy, University of California, Berkeley, CA 94720, USA}

\author{Melissa Soriano}
\affiliation{Jet Propulsion Laboratory, California Institute of Technology, 4800 Oak Grove Dr.,
  Pasadena, CA  91109, USA}

\author{Shinji Horiuchi}
\affiliation{CSIRO Astronomy \& Space Science/NASA Canberra Deep Space Communication Complex, P.{}O.~Box~1035, Tuggeranong ACT 2901, Australia}

\author{Jorge L.~Pineda}
\affiliation{Jet Propulsion Laboratory, California Institute of Technology, 4800 Oak Grove Dr.,
  Pasadena, CA  91109, USA}
  
\author{Tyrone McNichols}
\affiliation{California Institute of Technology, 1200 E California Blvd, Pasadena, CA 91125, USA}

\begin{abstract}

The Deep Space Network (DSN) enables NASA to communicate with its spacecraft in deep space. By virtue of its large antennas, the DSN can also be used as a powerful instrument for radio astronomy. Specifically, Deep Space Station (DSS) 43, the 70~m antenna at the Canberra Deep Space Communications Complex (CDSCC) has a K-band radio astronomy system covering a 10~GHz bandwidth at 17~GHz to 27~GHz. This spectral range covers a number of atomic and molecular lines, produced in a rich variety of interstellar gas conditions. Lines include hydrogen radio recombination lines (RRLs), cyclopropenylidene (C${}_3$H${}_2$), water masers (H${}_2$O), and ammonia (NH${}_3$). A new high-resolution spectrometer was deployed at CDSCC in November 2019 and connected to the K-band downconverter. The spectrometer has a total bandwidth of 16 GHz. Such a large total bandwidth enables, for example, the simultaneous observations of a large number of RRLs, which can be combined together to significantly improve the sensitivity of these observations. The system has two firmware modes: 1) A 65k-pt FFT to provide 32768 spectral channels at 30.5~kHz and 2) A 16k-pt polyphase filterbank (PFB) to provide 8192 spectral channels with 122~kHz resolution. The observation process is designed to maximize autonomy, from the Principle Investigator's inputs to the output data in FITS file format. We present preliminary mapping observations of hydrogen RRLs in Orion KL mapping taken using the new spectrometer. 

\end{abstract}

\section{Introduction} \label{sec:intro}

Spectroscopic observations in the radio part of the electromagnetic
spectrum provide key information in the physical and chemical state of
a variety of astrophysical environments. These environments include
the ionized medium created by massive stars, traced via observations
of hydrogen radio recombination lines (RRLs), the densest parts of
molecular clouds where star formation takes place, and protoplanetary
disks where planets form, both of which are traced by observations of
molecular line emission. All these science applications can be more efficient with the
capability to observe spectral lines over a large instantaneous
bandwidth. Spectral lines from RRLs and complex molecules are
dispersed over a large range of frequencies, and therefore stacking
techniques can be used for obtaining sensitive
observations. Additionally, broadband capabilities are required for
efficient spectral line surveys of star forming regions in the Milky
Way and nearby galaxies.

The spectral range of approximately 17~GHz to 27~GHz (K--band) covers
a number of atomic and molecular lines, produced in a rich variety of
interstellar gas conditions. Lines include hydrogen radio
recombination lines (RRLs), cyclopropenylidene (C${}_3$H${}_2$), water (H${}_2$O), and
ammonia (NH${}_3$). Also, redshifted spectral lines can be observed in K--band from distant galaxies,
e.g., silicon monoxide (SiO). In Table~\ref{tab:selectlines}, we present a selection of key molecular
and ionic species that can be observed in K--band.

In this paper, we present a new broadband spectrometer that covers 16 1-GHz IFs in the
17~GHz to 27~GHz range simultaneously, allowing efficient line surveys 
in galactic and extragalactic sources. This spectrometer operates in K--band and has been installed at the Deep Space Network (DSN) Complex
in Canberra, Australia.

In \S\ref{sec:architecture} we describe
the architecture of the spectrometer; in \S\ref{sec:data} we describe the data products; in
\S\ref{sec:implement} we describe the implementation and verification; and in
\S\ref{sec:conclude} we present our conclusions and possible paths
toward the future.

\begin{deluxetable}{lll} 
\tabletypesize{\scriptsize}
\centering \tablecolumns{3} \small
\tablecaption{ Selected Lines in the 17 to 27 GHz range$^1$.}
\label{tab:selectlines}
\tablehead{\colhead{Line Name} & \colhead{Frequency [GHz]} & \colhead{Maser or Thermal?$^2$}}
\startdata
H72$\alpha$          & 17.258          & Thermal \\
H71$\alpha$          & 17.992          & Thermal \\
NH${}_3$ (7,3)       & 18.017    & Thermal	   \\
NH${}_3$ (10,7)      & 18.285    & Thermal	   \\
NH${}_3$ (6,1)       & 18.391    & Thermal	   \\
H70$\alpha$          & 18.769          & Thermal \\
H69$\alpha$          & 19.591          & Thermal   \\
CH$_3$OH ($2_1-3_0$) & 19.967 & Both (II)	   \\
H68$\alpha$          & 20.462          & Thermal   \\
NH$_3$ (8,6)         & 20.719    & Both		   \\
NH$_3$ (9,7)         & 20.735    & Thermal	   \\
C6H ($15/2-13/2$)    & 20.792  & Thermal	   \\
NH${}_3$ (7,5)         & 20.804    & Thermal	   \\
NH$_3$ (11,9)        & 21.071   & Both		   \\
NH$_3$ (4,1)         & 21.134    & Thermal	   \\
H67$\alpha$          & 21.385     & Thermal	   \\
H$_2$O ($6_1-5_2$)   & 22.235   & Maser		   \\
CCS ($2{}_1-1_0$)      & 22.344      & Thermal	   \\
H66$\alpha$          & 22.364     & Thermal	   \\
HC$_7$N (20$\to$19)  & 22.559  & Thermal	   \\
NH$_3$ (2,1)         & 23.099    & Thermal	   \\
CH$_3$OH ($9{}_2-10{}_1$)     & 23.121 & Maser (II)	   \\
H65$\alpha$           & 23.404 & Thermal	   \\
CH$_3$OH ($10_1-9_2$) & 23.444 & Thermal	   \\
NH$_3$ (1,1) & 23.694 & Thermal			   \\
NH$_3$ (2,2) & 23.722 & Thermal			   \\
NH$_3$ (3,3) & 23.870 & Both			   \\
NH$_3$ (4,4) & 24.139 & Thermal			   \\
NH$_3$ (5,5) & 24.533 & Both			   \\
NH$_3$ (6,6) & 25.056 & Both			   \\
H64$\alpha$  &  24.509  & Thermal		   \\
CH$_3$OH ($3{}_2{}_3{}_1$) & 24.928 & Both (I)	   \\
CH$_3$OH ($4{}_2{}_4{}_1$) & 24.933 & Both (I)	   \\
CH$_3$OH ($2{}_2{}_2{}_1$) & 24.934 & Both (I)	   \\
CH$_3$OH ($5{}_2{}_5{}_1$) & 24.959 & Both (I)	   \\
CH$_3$OH ($6{}_2{}_6{}_1$) & 25.018 & Both (I)	   \\
CH$_3$OH ($7{}_2{}_7{}_1$) & 25.124 & Both (I)	   \\
H63$\alpha$   & 25.686 & Thermal		   \\
HC$_5$N (109) & 26.626 & Thermal		   \\
H62$\alpha$   & 26.939 & Thermal		   \\
\enddata 
\tablenotetext{1}{Adapted from Table~1 in \citet{Walsh2011}.}
\tablenotetext{2}{For CH$_3$OH masers, they are identified either as Class~I or~\hbox{II}.}
\end{deluxetable}

\section{Architecture}\label{sec:architecture}

\begin{figure*} 
\includegraphics[width=1\textwidth]{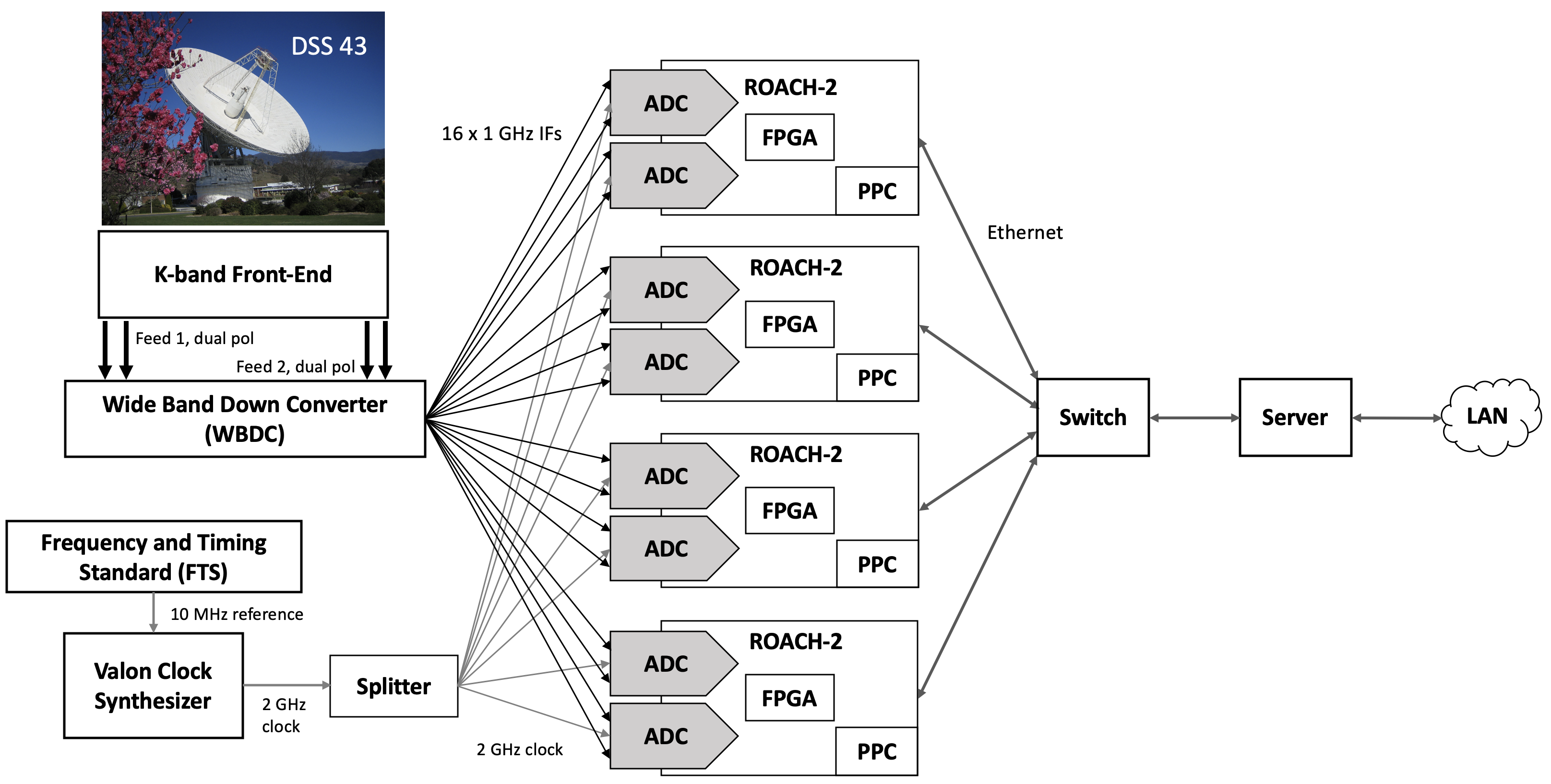}
\caption{Spectrometer architecture diagram showing the signal path from DSS-43 to the spectrometer.}
\label{fig:arch}
\end{figure*}
The 70~m antenna Deep Space Station 43 (DSS-43, also known as
``Tidbinbilla 70~m'') at the Canberra Deep Space Communications
Complex (CDSCC) within NASA's Deep Space Network is equipped with a K-band
radio astronomy system. Operating over the 17~GHz to 27~GHz frequency
range, the analog portion of the K-band system produces 40 intermediate frequencies
(IFs), each of 1~GHz bandwidth \citep{kfm+19}. These IFs from the Wide Band Down Converter (WBDC) at DSS-43 are distributed
to the new spectrometer. The full spectrometer can process up to 16 1-GHz IFs.

\begin{deluxetable}{ll}
\tablecaption{DSN Radio Astronomy spectrometer Modes}
\label{tab:specmodes} 
\tablewidth{0pc}
\tablehead{\colhead{Mode} & \colhead{Description}}
\startdata
1 & 32k spectral channels, FFT-based\\
& (30.5~kHz $\approx$ 0.4~km s$^{-1}$ at 22~GHz)\\
2 & 8k spectral channels, polyphase filterbank-based\\
& (122~kHz $\approx$ 1.7~km s$^{-1}$ at 22~GHz)\\
\enddata
\end{deluxetable}

As an initial implementation, two modes were designed ((Table~\ref{tab:specmodes}).
The focus was on providing a sufficient
spectral resolution for many forms of spectroscopy. 
Data from the Radio Astronomy spectrometer is provided in
single-dish FITS (SDFITS\footnote{The Registry of FITS Conventions,
\url{https://fits.gsfc.nasa.gov/fits\_registry.html}}) format \citep{g00}.
\subsection{Hardware Overview}
The spectrometer consists of four CASPER Reconfigurable Open Architecture Computing Hardware 2  \footnote{%
CASPER ROACH-2 boards,\url{https://casper.ssl.berkeley.edu/wiki/ROACH2}} boards as shown in Figure~\ref{fig:arch} (ROACH-2). 
Each board has 2 ADC (Analog to Digital Converters) cards, 
an \hbox{FPGA}, and PowerPC microprocessor. The ADC cards are CASPER 8 bit 5 GSPS ADC cards. Each ADC card has 2 inputs resulting in each board having a total of four inputs.

A Valon 5009 dual frequency synthesizer controls the 2~GHz clock for each of the ROACH-2 boards.
The Valon synthesizer receives a 10~MHz reference signal provided by the Frequency and Timing Standard (FTS) system available at \hbox{CDSCC}.

The ROACH-2 boards are controlled by a Dell PowerEdge R510 server. The server controls the boards by running Bash scripts which call Python software discussed in \ref{soft}. The server is connected to the DSN complex's Local Area Network (LAN) and the ROACH-2 boards are connected via a network switch to the server. The ROACH-2 boards form an internal network behind the switch. 

\subsection{Software Design} \label{soft}

\begin{figure*}
\includegraphics[width=1\textwidth]{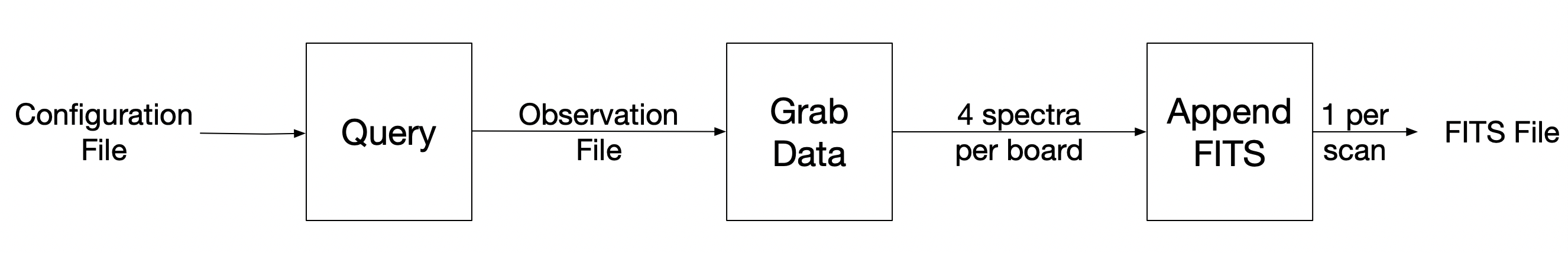}
\caption{Software process diagram shows 3 scripts: Query, Grab Data, and Append FITS which run during an observation.}
\label{figure:software}
\end{figure*}

The software design is shown in Figure~\ref{figure:software}. The
configuration file contains information such as the DSN antenna pointing filename, experiment stop
time, gap time between scans, observation mode, IF configuration, and system
temperature. Further detail on the information included in the configuration file and how the Principle Investigator (PI) supplies the input parameters can be
found in \S\ref{sec:datainput}. 

The Query box represents the query script, which creates
an observation file based on the schedule listed in the DSN antenna pointing file. The
times in the observation file and pointing file slightly differ since the observation
file takes into account antenna slew time between sources. 

The Grab Data box represents
the grab data script which controls the four ROACH-2 boards. The grab data script starts
and stops data acquisition on the ROACH-2 boards based on the observation file.  Data acquisition
begins at the specified start time in the observation file. Data acquisition ends when either the stop
time in the observation file is reached, or when the target number of samples is reached for a given
scan. Further detail on how the target number of samples is calculated can be found in
~\S\ref{sec:datainput}. 16 spectra data files are created for each scan. 

The Append FITS box represents the append
fits script. This script converts the binary spectra data results to FITS bintables.
The resulting FITS file includes 1 primary header and 16 bintables for each scan. 

\subsection{Firmware Design}
There are two firmware modes: a 65k-pt FFT with 32768 spectral channels with ~30.5 kHz resolution and a 16k-pt polyphase filterbank (PFB) with 8192 spectral channels with ~122 kHz resolution. The 8k mode scripts refer to the 16k-pt PFB and the 32k mode scripts refer to the 65k-pt FFT. The PI selects the mode for their track. For RRL mapping, the 65k-pt FFT mode is operationally used.

\begin{figure*}
\includegraphics[width=1\textwidth]{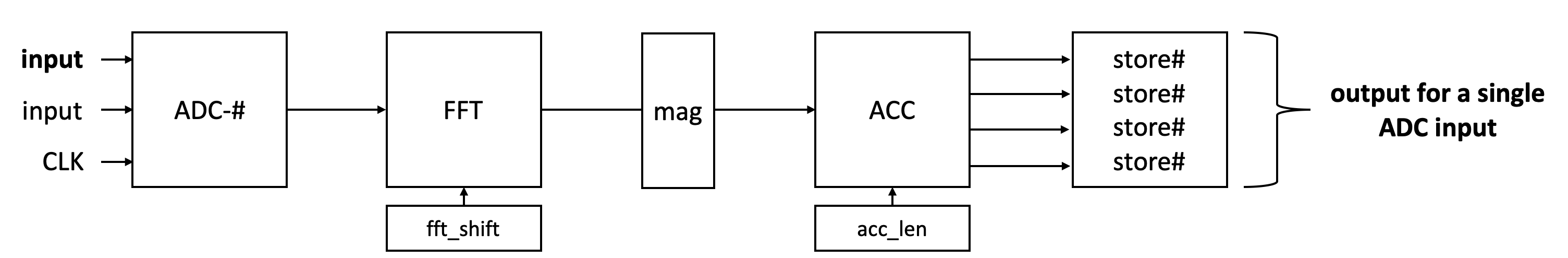}
\caption{Firmware block diagram showing process for a single input to a single ADC for both firmware modes}
\label{figure:firm}
\end{figure*}

A general firmware block diagram is shown in Figure~\ref{figure:firm} for both 8k and 32k modes.
The 32k mode is described as follows for a single input on a single ADC. The ADC data from a single
input is 65536 real samples. The ADC data is then put through an FFT. The FFT results in 32768  
complex channels. The complex magnitude is taken. The ACC block is the accumulator which averages
the data a certain number of times as specified by acc\_len.  

The accumulator integrates the data as 32 bit samples. Then the averaged FFT data is split
into four 8192 channel buffers which have 32 bits/channel. Each 8192 channel buffer is put in a
store register. For example, the data for one of the ADC inputs is stored in the first 4 store registers. The other ADC input stores its data in the next 4 store registers and so on. This firmware
process is duplicated 4 times, once for each of the 4 inputs on each ROACH-2 board.  
The 8k firmware mode process is similar. The ADC data from a single input is 16384 real samples.
After the PFB, there are 8192 complex channels. After integration, the data is split into four 2048 channel buffers.

\section{Data Products}\label{sec:data}

The spectrometer and associated software are designed to produce a set
of data products, consistent with the model used in many NASA
missions and radio observatories. There is a set of Level~0 data products that are
intended to form a self-contained set of data from which all
subsequent processing can proceed. Higher level data products are
generated, but in a manner that allows either for alternate processing
or modified processing.

\begin{deluxetable}{ll} 
\tabletypesize{\scriptsize}
\centering \tablecolumns{2} \small
\tablewidth{10cm}
\tablecaption{PI Input Information for an Observation} 
\label{tab:inputable}
\tablehead{\colhead{Item Name} & \colhead{Description}}
\startdata
Group Name & PI-specified group name\\
Source Name & PI-specified source name for each source \\
Right Ascension & Epoch J2000 (hh:mm:ss) \\
Declination & Epoch J2000 (sdd:mm:ss)\\
Spectral Resolution & 30.5kHz or 122kHz\\
Start Time & For all scans (yy:doy:hh:mm:ss)\\
Final Stop Time & Stop time for the final scan (yy:doy:hh:mm:ss)\\
Observation Mode & OTF or PSW\\
Sample Cadence & 2 options available for each spectral resolution\\
IF Selection & 16 IFs within 17GHz-27GHz dual-polarized IFs\\ 
\enddata 
\end{deluxetable}
\subsection{Input Parameters}\label{sec:datainput}
The PI provides the input parameters listed in Table ~\ref{tab:inputable}. The available coordinate system is equatorial. The right ascension and declination must be in the J2000 epoch. K-band in the range 17 GHz to 27 GHz is currently the only available band. However, the IF inputs to the spectrometer can be manually connected to the X-band or Ka-band downconverters if desired. 

The information from the PI is used to create the configuration file which is the input to the spectrometer's software. These input parameters are also used to generate the set of pointing files that are used within the DSN to point the antenna.

The sample
cadence or sample integration time is the time that the spectrometer accumulates for. This sample cadence can be calculated using the clock rate, number of samples and the acc\_len register value. The sample cadence calculations are shown in Equations \ref{eqn:samplecad} and \ref{eqn:samplecad2}.\\ 
\begin{equation}
number\_of\_samples / clock\_rate = time\_per\_sample\\
  \label{eqn:samplecad}
\end{equation}

\begin{equation}
acc\_len * time\_per\_sample = sample\_cadence
  \label{eqn:samplecad2}
\end{equation}
The sample cadence options for the 30.5~kHz spectral resolution are 4.29~seconds and 2.15~seconds.
The sample cadence options for the 122~kHz spectral resolution are 1.07~seconds and 2.15~seconds.

For a given scan, the target number of samples is calculated
 using the sample cadence, the duration of the scan, and a value called allowance. Allowance accounts for the time it takes for the script to read the data from the store registers. Introducing it ensures that the scan will not last longer than anticipated.
 For 32k mode, allowance is 1.75 seconds, and for 8k mode, it is 0.85 seconds. The calculation for the target number of samples is shown in Equation \ref{eqn:allow}.
 \begin{equation}
target\_number\_of\_samples = \frac{(duration - allowance)}{sample\_cadence}
 \label{eqn:allow}
 \end{equation}

\subsection{Level~0 Data}\label{sec:data0}

The Level~0 data are provided as SDFITS format files \citep{g00}. The primary
header contains relevant data about the antenna and details of the
observations, with Table \ref{tab:header} in Appendix~\ref{app:header} providing an illustration
of such a header. The observation modes are either on the fly mapping (OTF) or position switching (PSW). 
PSW is a single integration in a given position following with an integration in the emission--free region (``off'' position).
In the OTF mode \citep{Mangum2007, Wong2016}, the telescope moves at a certain rate while the spectrometer collects data followed by an observation on the ``off'' position, enabling efficient spectral line mapping. 
Additional data tables (structured as FITS \texttt{BINTABLE}s) are
provided, which contain the spectra data and information about the
observation that is used for subsequent processing to higher level 
data products, such as the system temperature for each sample, and the angular offset from the source coordinates for OTF mode. There are 16 bintables per scan, one for each input of the
spectrometer. The bintables are composed of 7 columns by $n$ rows.
The number of rows is determined by the number of samples in the 
spectrum. The sample cadence calculation is shown in \ref{sec:datainput}. If a sample is not complete before the observation
time ends, it is not included in the spectra. Each bintable has a header,  
Table \ref{tab:binheader} Appendix~\ref{app:header} provides an illustration
of such a header.
The 7 columns are as follows: spectrum data, antenna's elevation in meters as 
a function of time during each scan, an estimate of the system temperature~$T_{\mathrm{sys}}$ in Kelvins as a 
function of time during each scan, bandwidth in in Hz, duration in seconds, right ascension offset in degrees, and declination offset in degrees. The last two columns (right ascension offset and declination offset) are relevant for OTF observations.

\subsection{Level~1 Data}\label{sec:data1}

These data products are intended to be a notional processing of the
Level~0 data, in a manner that is consistent with producing data
products sufficient for analysis and publication but also for
alternate or additional processing. The intent is to provide a basic
estimate of the antenna temperature using the standard
position-switching technique,
\begin{equation}
T^*_{\mathrm{ant}} = \frac{T_{\mathrm{on}} - T_{\mathrm{off}}}{T_{\mathrm{off}}}T_{\mathrm{sys}},
  \label{eqn:anttemp}
\end{equation}
where $T_{\mathrm{on}}$ is the measured temperature (power) toward a
source, $T_{\mathrm{off}}$ is the measured temperature (power)
toward an ``off'' position, and $T_{\mathrm{sys}}$ is the system temperature.

The calibrated data is then written in the FITS format and, optionally, in the GILDAS/CLASS format \citep{Pety2005}, for further processing.

The Level~1a data product performs the first portion of the
calibration, forming the factor (${T_{\mathrm{on}} -
  T_{\mathrm{off}}})/{T_{\mathrm{off}}}$ using the
\emph{nearest-neighbor} ``off'' scan. This division in the
calibration is structured in this manner for a number of motivations.
With this initial ``ON$-$OFF'' calculation an investigator can assess the
calibration at this stage, there may be some cases in which only
relative calibration (without the system temperature) may be valuable,
and there are alternate approaches to identifying the ``off'' scan
(e.g., linear interpolation between the two ``off'' scans nearest in time).

The Level~1b data product uses the Level~1a data product and an
estimate of the system temperature~$T_{\mathrm{sys}}$ to form the
antenna temperature~$T^*_{\mathrm{ant}}$ of equation~(\ref{eqn:anttemp}).

\section{Implementation and Commissioning}\label{sec:implement}

The spectrometer was installed and commissioned in the signal processing center at the CDSCC in November of 2019. In the following we describe the steps taken to test the system during commissioning. 

\subsection{Testing}

The spectrometer has test software which provides real-time ADC and FFT plots. This software was used for commissioning and operations. ADC samples can be captured and an FFT can be computed in software.  
When the spectrometer arrived at CDSCC, both firmware modes were tested with K band inputs to check out the system after transport and ensure that no damage had occurred. The firmware accumulates a single sample and calculates an FFT, and this was also plotted and compared to the software computed FFT.  

On November 21, 2019 at 22:53 UTC, the spectrometer was also tested with dual-polarization X-band IFs as inputs. These operational bands are about 600 MHz wide and are typically used for spacecraft communications and tracking. The Juno spacecraft was being tracked at the time of the test and was emitting a carrier signal with a well known frequency. Although the X-band IFs do not exercise the full 1 GHz capability of each spectrometer input, this was a valuable test in validating the end to end system from a spacecraft all the way to the spectrometer processing.

\begin{figure}
\includegraphics[width=0.48\textwidth]{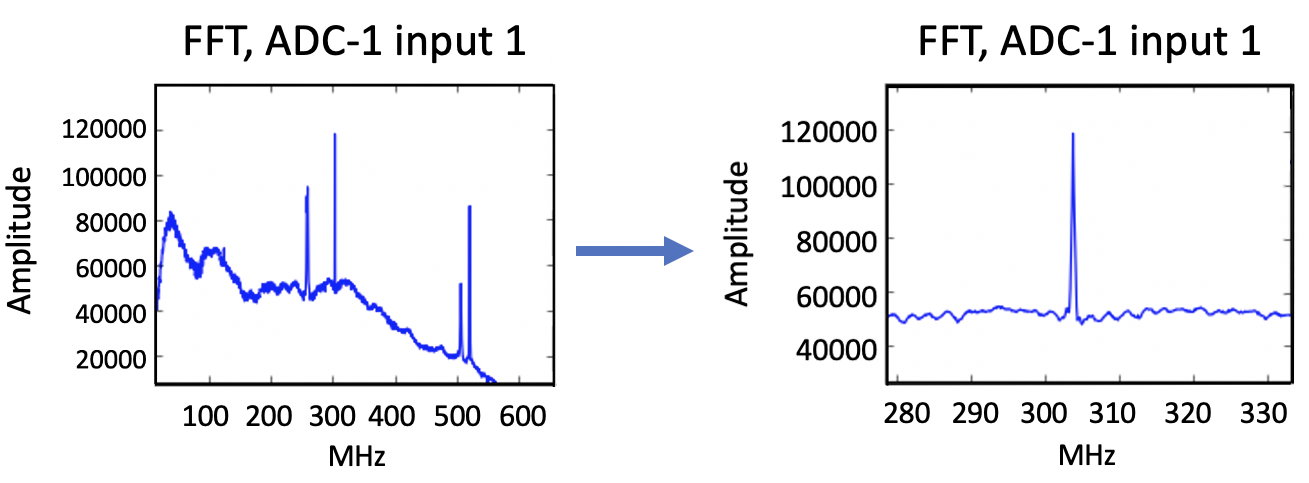}
\caption{On the left, plot of the down-converted Juno X-band carrier and on the right, a zoomed in plot of Juno X-band carrier at 303.75 MHz}
\label{figure:Jcarrierzoom}
\end{figure}

Figure \ref{figure:Jcarrierzoom} shows a single polarization of the X-band carrier from the Juno spacecraft during the test. The 0 to 600 MHz in the plot corresponds to 8100 to 8700 MHz. The signal is using the standard DSN X-band signal path for spacecraft communications, which are typically between 8200 and 8600 MHz and correspond to 100-500 MHz in Figure \ref{figure:Jcarrierzoom}. The X-band signal was down-converted by 8.1 GHz to an intermediate frequency of 303.75 MHz and then processed and recorded by the spectrometer. The right plot shows a zoomed-in version of the spectrum near the Juno carrier. This plot shows the spectrometer measurement matched the expected 303.75 MHz frequency.

\subsection{Verification}
\begin{figure*}
\includegraphics[width=1\textwidth]{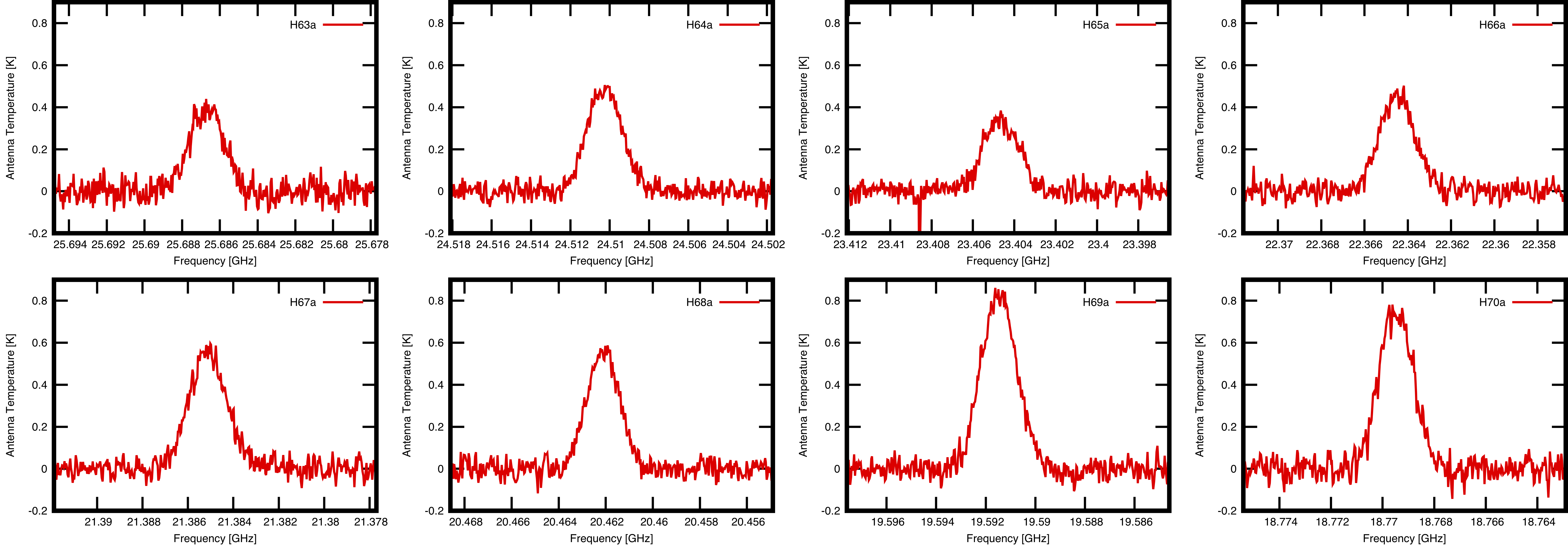}
\caption{18-26 GHz Hydrogen RRLs for Orion KL. The intensities obtained for the hydrogen recombination lines in Orion KL are consistent with those observed  by \citet{Gong2015} with Effelsberg 100m telescope.}
\label{figure:results}
\end{figure*}

In Figure\,\ref{figure:results}, we show a sample observation of hydrogen recombination lines in Orion KL. On November 24, 2019 we observed 8 RRLs simultaneously, from H70$\alpha$ to H63$\alpha$, each in two polarizations. The intensities obtained for the hydrogen recombination lines in Orion KL are consistent with those observed  by \citet{Gong2015} with Effelsberg 100m telescope, for a main--beam efficiency of 0.5 and  $S_{\nu}/T_{A}^{*}$= 1.9\,Jy/K \citep{kfm+19}. Note the difference in frequency response for each IF is caused by the Front End K-band system (see Figure 6 in \cite{kfm+19}). A powerful application for the spectrometer presented here is the possibility of stacking different RRLs, to significantly improve the signal--to--noise ratio of these observations, enabling the detection of faint lines, or for efficient OTF mapping over large spatial scales. 

\section{Conclusion}\label{sec:conclude}

A new high-resolution spectrometer was designed, implemented, and deployed to CDSCC in November 2019.  The spectrometer can process and record 16 1-GHz inputs simultaneously, providing a factor of 4 improvement in efficiency compared to the previous capability. This large total bandwidth, for example, enables the simultaneous observations of a large number of RRLs, which can be combined together to significantly improve the sensitivity of these observations.  The spectrometer software interfaces with the DSN and the PI to control observations and generate SDFITS files.

On November 21, 2019 at 22:53 UTC, the spectrometer was tested using the Juno spacecraft signal and the X-band system.  It was then connected to the K-band downconverter and commissioned by observing the 18-26 GHz hydrogen RRLs in Orion KL.  After commissioning was completed, an extensive mapping campaign of the Carina Nebula began.  Over 30 observations took place from November 2019 to March 2020. In March 2020, DSS-43 scheduled downtime began, which resulted in the K-band system also going offline. Observations are planned to begin again in early 2021.  The new high resolution spectrometer will enable future K-band observations by the DSN.  

\section{Future Work}\label{sec:futurework}

Additional modes could be implemented in the future.
For instance, a third mode providing an FFT-based 512M spectral point spectrum (1.9~Hz spectral resolution) has been investigated.

\acknowledgements

Part of this research was carried out at the Jet Propulsion
Laboratory, California Institute of Technology, under a contract with the National Aeronautics and Space Administration (80NM0018D0004).

\facilities{Deep Space Network}
\clearpage
\appendix

\section{DSN SDFITS Primary Header Example}\label{app:header}

Table~\ref{tab:header} and Table~\ref{tab:binheader} illustrate an example of an SDFITS file header
provided by the DSN Radio Astronomy spectrometer.

\begin{deluxetable}{ll}[h]
\tablecaption{Illustrative SDFITS Primary Header\label{tab:header}}
\tablewidth{0pc}
\tablecolumns{2}
\tablehead{\colhead{SDFITS \texttt{KEY}-\texttt{VALUE} pairs}&\colhead{Comment}}
\startdata
SIMPLE  =                   T & / conforms to FITS standard\\
BITPIX  =                    8 &/ array data type\\
NAXIS   =                    0 &/ number of array dimensions\\
EXTEND  =                    T\\
ORIGIN  = 'Python script'      &/ File origin\\
TELESCOP= 'DSS 43  '           &/ Observer\\
START   = '2020-069T11:14:10'  &/ Start time\\
STOP    = '2020-069T11:15:00'  &/ Stop time\\
DATE    = '2020 069'           &/ Date this file was written\\
SITELEV = '688.867 '           &/ Site elevation in meters\\
SITELAT = '-35     '           &/ Site latitude in degrees\\
SITELONG= '148     '           &/ Site longitude in degrees\\
DURATION= '00:00:10 '          &/ Duration of scan hh:mm:ss\\
SOURCE  = 'G305.1+0.0\_OFF'     &/ Object\\
RA      = '196.8749'           &/ Right ascension in degrees\\
DEC     = '-62.0600'           &/ Declination in degrees\\
INSTRUME= 'DSN RA SPECTROMETER' &/ Instrument\\
GUIDEVER= '1.0     '           &/ ROACH-2 User Guide version\\
FITSVER = '1.7     '           &/ FITS definition version\\
OBSMODE = 'PSW '           &/ Observation mode\\
COMMENT ROACH-2 DSN RA spectrometer\\
END\\
\enddata

\end{deluxetable}

\begin{deluxetable}{ll}
\tablecaption{Illustrative SDFITS Bintable Header\label{tab:binheader}}
\tablewidth{0pc}
\tablecolumns{2}
\tablehead{\colhead{SDFITS \texttt{KEY}-\texttt{VALUE} pairs}&\colhead{Comment}}
\startdata
XTENSION= 'BINTABLE'           &/ binary table extension\\
BITPIX  =                    8 &/ array data type\\
NAXIS   =                    2 &/ number of array dimensions\\
NAXIS1  =               131088 &/ length of dimension 1\\
NAXIS2  =                   2 &/ length of dimension 2\\
PCOUNT  =                    0 &/ number of group parameters\\
GCOUNT  =                    1 &/ number of groups\\
TFIELDS =                    7 &/ number of table fields\\
ORIGIN  = 'roach5-spec4'       &/ ROACH-2 board specification\\
CTYPE1  = 'FREQ    '           &/ CTYPE1\\
CRVAL1  =        21000000000.0 &/ Hz\\
CRPIX1  =                    0 &/ CRPIX1\\
CDELT1  =         30517.578125 &/ Hz\\
BUNIT   = 'K       '           &/ Kelvin\\
POL     = 'H       '           &/ Polarization\\
EXTNAME = 'SINGLE DISH'        &/ extension name\\
TTYPE1  = 'SPECTRUM'\\
TFORM1  = '32768E  '\\
TTYPE2  = 'TSYS    '\\
TFORM2  = 'E       '\\
TTYPE3  = 'ELEVATION'\\
TFORM3  = 'E       '\\
TTYPE4  = 'BANDWID '\\
TFORM4  = 'E       '\\
TTYPE5  = 'DURATION'\\
TFORM5  = 'E       '\\
TTYPE6  = 'RAOFFSET'\\
TFORM6  = 'E       '\\
TTYPE7  = 'DECOFFSET'\\
TFORM7  = 'E       '\\
COMMENT individual 1 GHz output from roach5-spec4\\
END\\
\enddata

\end{deluxetable}

\clearpage
\bibliography{papers.bib}{}
\bibliographystyle{aasjournal}

\end{document}